\documentclass[12pt]{iopart}
\usepackage{graphicx}
\usepackage{color}

\newcommand{\calE}{{\cal E}}

\newcommand{\calH}{{\cal H}}
\newcommand{\calI}{{\cal I}}

\newcommand{\calS}{{\cal S}}

\newcommand{\calU}{{\cal U}}
\newcommand{\calV}{{\cal V}}

\newcommand{\calX}{{\cal X}}

\begin{document}

\title[Work and Entropy Production in Information--Driven Engines]
{Relations Between Work and Entropy Production for General Information--Driven,
Finite--State Engines}

\author{Neri Merhav}

\address{The Andrew \& Erna Viterbi Faculty of Electrical Engineering, Technion, Haifa 32000,
Israel.\\ E--mail: merhav@ee.technion.ac.il}

\begin{abstract}
We consider a system model of a general finite--state machine (ratchet) that
simultaneously interacts with
three kinds of reservoirs: a heat reservoir, a work reservoir, and an information
reservoir, the latter being taken to be a running digital tape whose symbols interact
sequentially with the machine.
As has been shown in earlier work, this finite--state machine can
act as a demon (with memory), which creates a net flow of energy from the
heat reservoir into the work reservoir (thus extracting useful work) at the
price of increasing the entropy of the information reservoir.
Under very few assumptions, we propose a simple derivation of a family of inequalities that relate the
work extraction with the entropy production. These inequalities can be seen as
either upper bounds on the extractable work or as lower bounds on the entropy
production, depending on the point of view. Many of these bounds are relatively
easy to calculate and they are tight in the sense that equality 
can be approached arbitrarily closely.
In their basic forms, these inequalities are applicable to any finite number
of cycles (and not only asymptotically), and for a general input information
sequence (possibly correlated), which is not necessarily assumed even stationary. Several known
results are obtained as special cases.
\end{abstract}

\maketitle

\section{Introduction}

The fact that information processing plays a very interesting role in thermodynamics,
has already been recognized
in the second half of the nineteenth century, 
namely, when Maxwell proposed his celebrated gedanken experiment,
known as Maxwell's demon \cite{maxwelldemon}. According to the Maxwell demon
experiment, a demon with access to information on momenta and positions of particles
in a gas, at every given time, is cable of separating between fast--moving particles and slower ones, thus
forming a temperature difference without supplying external energy,
which sounds in contradiction to the second law of thermodynamics.
A few decades later, Szilard \cite{szilardeng} 
pointed out that it is possible to convert heat into work, when considering
a box with a
single particle. In particular, using a certain protocol of measurement and control, one may
be able to produce work in each cycle of the system, which is again, in
apparent contradiction with to the second law, since no external energy is
injected. 

These intriguing observations have
created a considerable dispute and controversy in the scientific community.
Several additional thought--provoking gedanken experiments have
ultimately formed the basis
for a vast amount of theoretical work associated with the role 
of informational ingredients in thermodynamics. 
An incomplete 
list of modern articles along these lines, include
\cite{AJM09}, \cite{BS13}, \cite{BS14a}, \cite{BS14b}, \cite{BBS14},
\cite{BMC16a}, \cite{BMC16b}, \cite{BMC16c},
\cite{CGQ15}, \cite{Deffner13}, \cite{DJ13}, \cite{EV11}, \cite{GDC13},
\cite{HBS14}, \cite{HA14}, \cite{HE14}, \cite{HS14}, \cite{MJ12}, \cite{MQJ13},
\cite{Merhav15}, \cite{PHS15}, \cite{SU12},
and \cite{SU13}.
These articles can be basically divided into two main categories. In the first
category,
the informational ingredient is in the form of measurement and
feedback control (just like in the Maxwell's demon and Szilard's engine) and
the second category is about
physical systems that include, beyond the
traditional heat reservoir (heat bath), also a work reservoir and an {\it information
reservoir}, which interacts with the system entropically, but with no energy
exchange. The information reservoir, which is a relatively new concept in physics \cite{BS14a},
\cite{BS14b}, \cite{DJ13},
may be, for instance, a large memory register or
a digital tape carrying a long sequence of bits, which interact sequentially with
the system and may change during this interaction.
Basically, the main results, in all these articles, are generalized forms
of the second law of thermodynamics, where the 
entropy increase consists of an 
extra term that is concerned with information
exchange, such as mutual information (for systems with measurement and feedback control) or
Shannon entropy increase (for systems with a information reservoir).

In contrast to the early proposed thought experiments, that were typically described in
general terms of an ``intelligent agent'' and were not quite described in full detail,
Mandal and Jarzynski \cite{MJ12} were the first to devise a concrete model of a
system that behaves basically like a demon. Specifically, they
described and analyzed a simple autonomous system, based on 
a finite--state Markov process,
that when operates as an engine, it converts
heat into mechanical work, and, at the same time, it writes
bits serially on a 
tape, which plays the role of an information reservoir.
Here, the word ``writes'' refers to a situation where the entropy of the output
bits recorded on the tape (after the interaction), is larger than the entropy of the input bits
(before the interaction).
It can also act as an eraser, which performs
the reversed process of losing energy while ``deleting'' information, that is,
decreasing the entropy.
Several variants on this
physical model, which are based on quite similar ideas, were offered in some
later articles. These include:
\cite{BS13} -- where the running tape can move both
back and forth,
\cite{BS14a} -- where the interaction time with 
each bit is a random variable rather than fixed parameter, 
\cite{BS14b} -- with three different points of view on information--driven systems,
\cite{BBS14} -- with the upper energy level being time--varying,
\cite{CGQ15} -- with a model based on
enzyme kinetics,
\cite{Deffner13} -- with a quantum model, 
\cite{HA14} -- with a thermal
tape, and 
\cite{MQJ13}, which concerns an information--driven refrigerator,
where instead of the work, heat is transferred from a cold reservoir
into a hotter one.

In a recent series of interesting papers, \cite{BMC16a}, \cite{BMC16b}, \cite{BMC16c},
Boyd, Mandal and Crutchfield considered a system model of a demon (ratchet)
that is implemented by a general finite--state machine (FSM) that simultaneously interacts with a heat
reservoir (heat bath at fixed temperature), 
a work reservoir (i.e., a given mass that may be lifted by the machine), and an information reservoir 
(a digital tape, as described above). The state variable of the FSM, which
manifests the memory of the ratchet to past input and output information,
interacts with the current bit of the information reservoir during one unit of
time, a.k.a.\ the interaction interval (or cycle), and then the machine produces the
next state and the output bit, before it turns to process the next input bit,
etc. The operation of the ratchet during one cycle
is then characterized by the joint probability distribution of
the next state and the output bit given the current state and the input bit. Perhaps
the most important result in \cite{BMC16a}, \cite{BMC16b} and \cite{BMC16c},
is that for a stationary input process (i.e., the incoming sequence of tape bits), 
the work extraction per cycle is asymptotically upper bounded by $kT$ times the difference
between the Shannon entropy rate of the tape output process and that of the 
input process (both in units of nats\footnote{$1~\mbox{nat}=\log_2\mbox{e}$
bits. Entropy defined using the natural base logarithm has units of nats.}
per cycle), i.e., eq.\ (5) of \cite{BMC16a} (here $k$ is the Boltzmann
constant and $T$ is the temperature). In addition to this general result,
various conclusions are drawn in those papers. For example, 
the uselessness of ratchet memory when
the input process is memoryless (i.i.d.), as well as its usefulness (for maximizing
work extraction) when the input
process is correlated, are both discussed in depth, and several interesting examples are demonstrated.
While the above mentioned upper bound on the work extraction, \cite[eq.\
(5)]{BMC16a}, seems reasonable and interesting, some concerns arise upon
reading its derivation in \cite[Appendix A]{BMC16a}, and these concerns are discussed
in some detail in the Appendix. 

In this paper, we consider a similar setup, but we focus is on the derivation of a family
of alternative inequalities that relate work extraction to entropy production.
The new proposed inequalities have the following advantages.
\begin{enumerate}
\item The approach taken and the derivation are very simple. 
\item The underlying assumptions about the input process, the ratchet,
and the other parts of the system, are rather mild.
\item The inequalities apply to any finite number of cycles.
\item For a stationary input process, the inequalities are simple
and the resulting bounds are relatively easy to calculate.
\item The inequalities are tight in the sense that equality can be approached
arbitrarily closely. 
\item Some known results are obtained as special cases.
\end{enumerate}

The remaining part of the paper is organized as follows.
In Section 2, we establish some notation conventions.
In Section 3, we describe the physical system model.
In Section 4, we derive
our basic work/entropy--production inequality.
In Section 5, we discuss this inequality and explore it from various points of
view. Finally, in Section 6, we derive a more general family of inequalities,
which have the flavor of fluctuation theorems.

\section{Notation Conventions}

Throughout the paper, random variables will be denoted by capital
letters, specific values they may take will be denoted by the
corresponding lower case letters, and their alphabets
will be denoted by calligraphic letters. Random
vectors, their realizations and their alphabets will be denoted,
respectively, by capital letters, the corresponding lower case letters,
and the corresponding calligraphic letters,
all superscripted by their dimension.
For example, the random vector $X^n=(X_1,\ldots,X_n)$, ($n$ --
positive integer) may take a specific vector value $x^n=(x_1,\ldots,x_n)$
in $\calX^n$, which is the $n$--th order Cartesian power of $\calX$,
the alphabet of each component of this vector.
The probability of an event $\calE$ will be denoted by $P[\calE]$.
The indicator function of an event $\calE$ will be denoted by $\calI[\calE]$.

The Shannon entropy of a discrete random 
variable $X$ will be denoted\footnote{Following the
customary notation conventions in information theory, $H(X)$ should not be
understood as a function $H$ of the random outcome of $X$, but as a functional
of the probability distribution of $X$.} by $H(X)$, that is,
\begin{equation}
\label{entropydef}
H(X)=-\sum_{x\in\calX}P(x)\ln P(x),
\end{equation}
where $\{P(x),~x\in\calX\}$ is the probability distribution of $X$. 
When we wish to emphasize the dependence of the entropy on the underlying
distribution $P$, we denote it by $\calH(P)$. The binary entropy function will
be defined as
\begin{equation} 
h(p)=-p\ln p-(1-p)\ln(1-p),~~~~0\le p\le 1.
\end{equation} 
Similarly, for a discrete random vector $X^n=(X_1,\ldots,X_n)$, the joint entropy
is denoted by $H(X^n)$ (or by $H(X_1,\ldots,X_n)$), and defined as
\begin{equation}
\label{jointentropydef}
H(X^n)=-\sum_{x^n\in\calX^n}P(x^n)\ln P(x^n).
\end{equation}
The conditional entropy of a generic random variable $U$ over a discrete
alphabet $\calU$,
given another generic random variable
$V\in\calV$, is defined as
\begin{equation}
\label{condentropydef1}
H(U|V)=-\sum_{u\in\calU}\sum_{v\in\calV}P(u,v)\ln P(u|v), 
\end{equation}
which should not be confused with the conditional entropy given a {\it specific
realization} of $V$, i.e.,
\begin{equation}
\label{condentropydef2}
H(U|V=v)=-\sum_{u\in\calU}P(u|v)\ln P(u|v). 
\end{equation}
The mutual information between $U$ and $V$ is
\begin{eqnarray}
I(U;V)&=&H(U)-H(U|V)\nonumber\\
&=&H(V)-H(V|U)\nonumber\\
&=&H(U)+H(V)-H(U,V),
\end{eqnarray}
where it should be kept in mind that in all three definitions, $U$ and $V$ can
themselves be
random vectors. Similarly,
the conditional mutual information between $U$ and $V$ given $W$ is
\begin{eqnarray}
I(U;V|W)&=&H(U|W)-H(U|V,W)\nonumber\\
&=&H(V|W)-H(V|U,W)\nonumber\\
&=&H(U|W)+H(V|W)-H(U,V|W).
\end{eqnarray}
The Kullback--Leibler divergence (a.k.a.\ relative entropy or
cross-entropy) 
between two distributions $P$ and $Q$ on
the same alphabet $\calX$, is defined as
\begin{equation}
D(P\|Q)=\sum_{x\in\calX}P(x)\ln\frac{P(x)}{Q(x)}.
\end{equation}

\section{System Model Description}

As in the previous articles on models of physical systems with an information reservoir,
our system consists of the following ingredients: a heat bath at temperature
$T$, a work reservoir, here designated by a wheel loaded by a mass $m$,
an information reservoir in the form of a digital input tape, a
corresponding output tape, and a certain device, which is the demon, or ratchet, in the
terminology of \cite{BMC16a}, \cite{BMC16b}, \cite{BMC16c}. The ratchet
interacts (separately)
with each one of the other parts of the system (see Fig.\ 1). 

\begin{figure}[ht]
\hspace*{3cm}\input{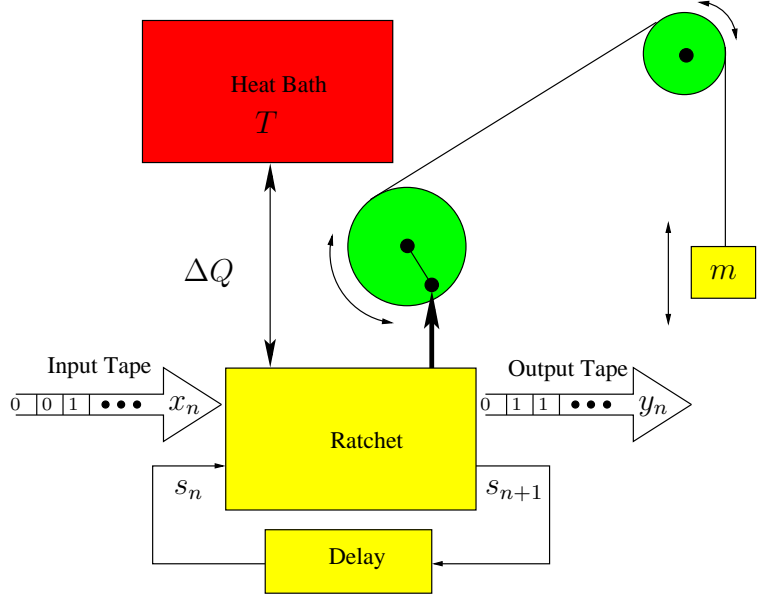}
\caption{The physical system model.}
\end{figure}

The input tape consists of a sequence of symbols, $x_1,x_2,\ldots$, from a
finite alphabet $\calX$ (say, binary symbols where $\calX=\{0,1\}$), that are
serially fed into the ratchet, which in turn processes these symbols
sequentially, while going through a sequence of internal states, $s_1, s_2, \ldots$, 
taking values in a finite set $\calS$. The ratchet outputs another sequence of
symbols, $y_1, y_2,\ldots$, which are elements of the same alphabet, $\calX$, as the
input symbols. The state of the ratchet is an
internal variable
that encodes the memory that the ratchet has with
regard to its history.
In the $n$--th cycle of the process ($n =
1,2,\ldots$), while the ratchet is at state $s_n$, it is fed by the input
symbol $x_n$ and it produces the pair 
$(y_n,s_{n+1})$ in stochastic manner, according to a given
conditional distribution, $P(y_n,s_{n+1}|x_n,s_n)$,
where $y_n$ is the output symbol at the $n$--th cycle and $s_{n+1}$ is the
next state. 

We now describe the mechanism that dictates this conditional
distribution, along with the concurrent interactions among the ratchet, the
heat bath and the work reservoir. The $n$--th cycle of the process occurs
during the time interval, $(n-1)\tau\le t < n\tau$, in other words, the
duration of each cycle is $\tau$ seconds, where $\tau > 0$ is a given
parameter. During each such interval, the
symbol and the state form together a Markov jump process, $(\xi_t,\sigma_t)$, whose 
state\footnote{Note that from this point and onward, there are two different
notions of ``state'', one of which is the state of ratchet, which is just $s_n$
(or $\sigma_t$), and the other one is the state of the Markov process, which
is the pair $(x_n,s_n)$ (or $(\xi_t,\sigma_t)$). To avoid confusion, we will
use the terms ``ratchet state'' and ``Markov state'' correspondingly, whenever
there is room for ambiguity.}
set is the product set $\calX\times\calS$ and whose matrix of Markov--state transition rates
is $M[(\xi,\sigma)\to(\xi^\prime,\sigma^\prime)]$, $\xi,\xi^\prime\in\calX$,
$\sigma,\sigma^\prime\in\calS$. 
The random Markov--state transitions of this
process are caused by spontaneous thermal fluctuations that result from the
interaction with the heat bath.
The Markov process is initialized at
time $t=(n-1)\tau$ according to 
$(\xi_{(n-1)\tau},\sigma_{(n-1)\tau})=(x_n,s_n)$.
At the end of this interaction interval, i.e., at time $t=n\tau - 0$, 
when the process is its final state $(\xi_{n\tau-0},\sigma_{n\tau-0})$,
the ratchet records the output symbol as $y_n=\xi_{n\tau-0}$ and the next
ratchet state becomes $s_{n+1}=\sigma_{n\tau-0}$, 
and then the $(n+1)$--st cycle
begins in the same manner, etc. 

Denoting
by $\Pi_t(\xi,\sigma)$ the probability of finding the Markov process in state
$(\xi,\sigma)$ at time $t$, it is clear from the above description, that
the conditional distribution
$P(y_n,s_{n+1}|x_n,s_n)$, that was mentioned before, is the solution 
$\{\Pi_{n\tau-0}(y,s)\}$ to the
master equations (see, e.g., \cite[Chap.\ 5]{vanKampen}),
$$\frac{\mbox{d}\Pi_t(\xi,\sigma)}{\mbox{d}t}=\sum_{\xi^\prime,\sigma^\prime}
\{\Pi_t(\xi^\prime,\sigma^\prime)M[(\xi^\prime,\sigma^\prime)\to(\xi,\sigma)]
-\Pi_t(\xi,\sigma)M[(\xi,\sigma)\to(\xi^\prime,\sigma^\prime)]\},$$
when the initial condition is $\Pi_{(n-1)\tau}(\xi,\sigma)=\calI\{(\xi,\sigma)=
(x_n,s_n)\}$.

Associated with each state, $(\xi,\sigma)$, of
the Markov process, there is a given energy $E(\xi,\sigma)=mg\cdot
\Delta(\xi,\sigma)$, 
$\Delta(\xi,\sigma)$ being the height level of the mass $m$ (relative to
some reference height associated with an arbitrary Markov state).
As the Markov
process jumps from $(\xi,\sigma)$ to $(\xi^\prime,\sigma^\prime)$, the ratchet
lifts the mass by $\Delta(\xi^\prime,\sigma^\prime)-\Delta(\xi,\sigma)$,
thus performing an amount of work given by $E(\xi^\prime,\sigma^\prime)-E(\xi,\sigma)$,
whose origin is heat extracted from the heat bath
(of course, the direction of the flow of energy between the heat bath and
the work reservoir is reversed when these energy
differences change their sign).
It should be pointed out that the input tape does not supply energy to the
ratchet, in other words, at the switching times, $t=n\tau$,
although the state of the Markov process changes from
$(\xi_{n\tau-0},\sigma_{n\tau-0})=(y_n,s_{n+1})$ to 
$(\xi_{n\tau},\sigma_{n\tau})=(x_{n+1},s_{n+1})$, this switching is not
assumed to be accompanied by a change in energy (the mass is neither raised nor
lowered). In other words, the various energy levels, $E(\\xi,\sigma)$, have only a relative
meaning, and so, after $N$
cycles, the total amount of work carried out by the ratchet is given by
\begin{equation}
W_N=\sum_{n=1}^N[E(y_n,s_{n+1})-E(x_n,s_n)].
\end{equation}

It will be assumed that the sequence of input symbols is governed by a
stochastic process, which is 
designated by $X_1,X_2,\ldots$, and which obeys a given probability law $P$, that is,
\begin{equation}
\mbox{Pr}\{X_1=x_1,~X_2=x_2,\ldots, X_n=x_n\}=P(x_1,x_2,\ldots,x_n),
\end{equation}
for every positive integer $n$ and every $(x_1,x_2,\ldots,x_n)\in\calX^n$,
where $P(x_1,x_2,\ldots,x_n)$ is the probability distribution function.
No special assumptions will be made concerning the process (not even
stationarity) unless this will be specified explicitly. 
Following the notation conventions described 
in Section 2, the notation of the input sequence using capital
$X$ emphasizes that this is a random process.
By the same token, when we wish to
emphasize the induced randomness of the ratchet state sequence and the output
sequence, we denote them by $\{S_n\}$ and $\{Y_n\}$, respectively.

To summarize, our model consists of two sets of stochastic processes in two different
levels: one level lies in the larger time scale which is discrete (indexed by
the integer $n$), and this is where the processes
$\{X_n\}$, $\{Y_n\}$ and $\{S_n\}$ take place.
The probability distributions of these
processes are denoted by the letter $P$. 
The other level is in the smaller time scale, which
is continuous, and this is where the Markov--jump pair process $\{(\xi_t,\sigma_t)\}$ takes place
during each interaction interval of length $\tau$. The joint probability
distribution of $(\xi_t,\sigma_t)$ is denoted by $\Pi_t$. The connection
between the two kinds of processes is that at times $t=(n-1)\tau$,
$n=1,2,\ldots$,
$(\xi_t,\sigma_t)$ is set to $(X_n,S_n)$, and at times $t=n\tau-0$,
$(Y_n,S_{n+1})$ is set to $(\xi_t,\sigma_t)$.

\section{The Basic Work/Entropy--Production Inequality}

As said, we are assuming that within each interaction interval, $(n-1)\tau\le t <
n\tau$, the pair $(\xi_t,\sigma_t)$ is a Markov jump process. 
For convenience of the exposition, let us temporarily shift the origin and
redefine this time interval
to be $0\le t < \tau$.
Since each Markov 
state $(\xi,\sigma)$, is associated with energy level $E(\xi,\sigma)$,
the equilibrium distribution is the canonical distribution,
\begin{equation}
\Pi_{\mbox{\tiny eq}}(\xi,\sigma)=\frac{e^{-\beta E(\xi,\sigma)}}{Z(\beta)},
\end{equation}
where $\beta=\frac{1}{kT}$ is the inverse temperature and
\begin{equation}
Z(\beta)=\sum_{(\xi,\sigma)\in\calX\times\calS} e^{-\beta E(\xi,\sigma)}.
\end{equation}
The Markovity of the
process implies 
that $D(\Pi_t\|\Pi_{\mbox{\tiny
eq}})$ is monotonically non--increasing
in $t$ (see, e.g., \cite[Chap.\ V.5]{vanKampen},
\cite[Theorem 1.6]{Kelly79}, \cite[Section 4.4]{CT06}),
and so,
\begin{equation}
\label{htheorem}
D(\Pi_\tau\|\Pi_{\mbox{\tiny eq}})\le
D(\Pi_0\|\Pi_{\mbox{\tiny eq}}),
\end{equation}
which is clearly equivalent to
\begin{equation}
\label{div}
\sum_{(\xi,\sigma)\in\calX\times\calS}
[\Pi_{\tau}(\xi,\sigma)-\Pi_0(\xi,\sigma)]\cdot
\ln\frac{1}{\Pi_{\mbox{\tiny eq}}(\xi,\sigma)}\le
\calH(\Pi_\tau)-\calH(\Pi_0).
\end{equation}
Since 
\begin{equation}
\label{p2w}
\ln\frac{1}{\Pi_{\mbox{\tiny
eq}}(\xi,\sigma)}=\ln Z(\beta)+\beta E(\xi,\sigma)
\equiv\ln Z(\beta)+\beta mg\Delta(\xi,\sigma),
\end{equation}
the left--hand side (l.h.s.) of (\ref{div}) gives the average work per
cycle (in units of $kT$), and the right--hand side (r.h.s.) is the difference between the
entropy of the final Markov state within the cycle, $(\xi_\tau,\sigma_\tau)$, and the 
entropy of the initial Markov state, $(\xi_0,\sigma_0)$. Returning to the
notation of the discrete time processes (indexed by $n$), we have then just shown
that
$$\left<\Delta W_n\right>\equiv
\left<E(Y_n,S_{n+1})\right>-\left<E(X_n,S_n)\right>\le
kT\cdot[H(Y_n,S_{n+1})-H(X_n,S_n)],$$
and so, the total average work after $N$ cycles is upper bounded by
\begin{equation}
\label{basic}
\left<W_N\right>\equiv \sum_{n=1}^N\left<\Delta W_n\right>\le
kT\cdot\sum_{n=1}^N [H(Y_n,S_{n+1})-H(X_n,S_n)]. 
\end{equation}
Eq.\ (\ref{basic}) serves as our basic work/entropy--production inequality.

A slightly different form is 
the following:
\begin{eqnarray}
\label{alternative}
\frac{\left<W_N\right>}{kT}&\le&
\sum_{n=1}^N [H(Y_n|S_{n+1})-H(X_n|S_n)]+ 
\sum_{n=1}^N [H(S_{n+1})-H(S_n)]\nonumber\\
&=&\sum_{n=1}^N [H(Y_n|S_{n+1})-H(X_n|S_n)]+ 
H(S_{N+1})-H(S_1).
\end{eqnarray}
The first sum in the last expression is the (conditional) entropy production associated with the input--output
relation of the system, whereas the
term $H(S_{N+1})-H(S_1)$ can be understood as the contribution of the
ratchet state to the net entropy production throughout the entire process of
$N$ cycles. 
If the ratchet has many states and $N$ is not too large, the latter contribution might be
significant, but if the number of ratchet states, $|\calS|$, is fixed, then
the relative contribution of ratchet--state entropy production term, which cannot
exceed $\ln|\calS|$, becomes negligible
compared to the input--output entropy production term for large $N$.
In particular, if we divide both sides of the inequality by $N$, then as
$N\to\infty$, the term
$\frac{\ln|\calS|}{N}$ tends to zero, and so, the the average work per
cycle is asymptotically upper bounded by $\frac{kT}{N}\sum_{n=1}^N
[H(Y_n|S_{n+1})-H(X_n|S_n)]$. This expression is different from the general
upper bound of
\cite{BMC16a},
\cite{BMC16b},
\cite{BMC16c}, where it was argued that $\left<W_N\right>/NkT$ is asymptotically
upper bounded by 
\begin{equation}
\label{bmc}
\frac{1}{N}[H(Y^N)-H(X^N)]=
\frac{1}{N}\sum_{n=1}^N[H(Y_n|Y^{n-1})-H(X_n|X^{n-1})].
\end{equation}
While both the first term in (\ref{alternative}) and (\ref{bmc}) involve sums of differences
between conditional output and input entropies, the conditionings
being used in the two bounds are substantially different. Our bound suggests that the relevant information
``memorized'' by both the input process and the output process, is simply the ratchet state
that is coupled to it, rather than its own past, as in (\ref{bmc}).
These conditionings on the states can be understood to be 
the residual input--output entropy production
that is {\it not} part of the entropy production of the ratchet state (which is in
general, correlated to the input and output).
Moreover, the last line of (\ref{alternative}) is typically easier to
calculate than (\ref{bmc}), as will be discussed and demonstrated in the
sequel.

Yet another variant of (\ref{basic}) is obtained when the chain rule of the
entropy is applied in the opposite manner, i.e.,
\begin{equation}
\frac{\left<W_N\right>}{kT}\le
\sum_{n=1}^N [H(Y_n)-H(X_n)]+ 
\sum_{n=1}^N [H(S_{n+1}|Y_n)-H(S_n|X_n)].
\end{equation}
Here the first term is the input--output entropy production and the second
term is the conditional entropy production of the ratchet state. However, this
form is less useful than (\ref{alternative}).

\section{Discussion on the Bounds and Their Variants}

In this section, we discuss eqs.\ (\ref{basic}) and (\ref{alternative}) as
well as several additional variants
of these inequalities.

\subsection{Tightness and Achievability}

The first important point concerning inequality (\ref{basic}) is that it is potentially tight in the sense
that the ratio between the two sides of eq.\ (\ref{basic}) may approach unity
arbitrarily closely. To see this, consider first the case where $\Pi_0(\xi,\sigma)$ is
close to $\Pi_{\mbox{\tiny eq}}(\xi,\sigma)$ in the sense that 
\begin{equation}
\Pi_0(\xi,\sigma)=\Pi_{\mbox{\tiny
eq}}(\xi,\sigma)\cdot[1+\epsilon(\xi,\sigma)],~~~(\xi,\sigma)\in\calX\times\calS
\end{equation}
where $\epsilon\equiv\max_{\xi,\sigma}|\epsilon(\xi,\sigma)| \ll 1$ and
obviously,
\begin{equation}
\label{avgeps}
\sum_{\xi,\sigma}\Pi_{\mbox{\tiny eq}}(\xi,\sigma)\epsilon(\xi,\sigma)=0
\end{equation}
since $\{\Pi_0(\xi,\sigma)\}$ must sum up to unity.
Assume also that $\Pi_\tau(\xi,\sigma)$ is even much closer to $\Pi_{\mbox{\tiny
eq}}(\xi,\sigma)$ in the sense that the ratio
$\Pi_\tau(\xi,\sigma)/\Pi_{\mbox{\tiny eq}}(\xi,\sigma)$ is between $1-\epsilon^2$
and $1+\epsilon^2$.
Now, the work per cycle is given by
\begin{eqnarray}
\left<\Delta W\right>&=&\sum_{\xi,\sigma}\Pi_\tau(\xi,\sigma)E(\xi,\sigma)-
\sum_{\xi,\sigma}\Pi_0(\xi,\sigma)E(\xi,\sigma)\nonumber\\
&=& \sum_{\xi,\sigma}\Pi_{\mbox{\tiny
eq}}(\xi,\sigma)E(\xi,\sigma)+O(\epsilon^2)-
\sum_{\xi,\sigma}\Pi_{\mbox{\tiny
eq}}(\xi,\sigma)[1+\epsilon(\xi,\sigma)]E(\xi,\sigma)\nonumber\\
&=&-\sum_{\xi,\sigma}\Pi_{\mbox{\tiny
eq}}(\xi,\sigma)\epsilon(\xi,\sigma)E(\xi,\sigma)+O(\epsilon^2).
\end{eqnarray}
On the other hand, the entropy production per cycle is given by
\begin{eqnarray}
\Delta\calH&\equiv&\calH(\Pi_\tau)-\calH(\Pi_0)\\
&=& \sum_{\xi,\sigma}\Pi_0(\xi,\sigma)\ln \Pi_0(\xi,\sigma)-
\sum_{\xi,\sigma}\Pi_\tau(\xi,\sigma)\ln \Pi_\tau(\xi,\sigma)\\
&=& \sum_{\xi,\sigma}\Pi_{\mbox{\tiny eq}}(\xi,\sigma)[1+\epsilon(\xi,\sigma)]
\ln\{\Pi_{\mbox{\tiny eq}}(\xi,\sigma)[1+\epsilon(\xi,\sigma)]\}-\nonumber\\
& &\sum_{\xi,\sigma}\Pi_{\mbox{\tiny eq}}(\xi,\sigma)\ln \Pi_{\mbox{\tiny
eq}}(\xi,\sigma)+O(\epsilon^2)\\
&=&\sum_{\xi,\sigma}\Pi_{\mbox{\tiny eq}}(\xi,\sigma)\epsilon(\xi,\sigma)\ln
\Pi_{\mbox{\tiny eq}}(\xi,\sigma)+O(\epsilon^2),
\end{eqnarray}
where the last line is obtained using (\ref{avgeps}).
Now, the difference $kT\Delta\calH-\left<\Delta W\right>$ is given
by $kT\cdot[D(\Pi_0\|\Pi_{\mbox{\tiny eq}})-D(\Pi_\tau\|\Pi_{\mbox{\tiny eq}})]$. But,
\begin{eqnarray}
D(\Pi_0\|\Pi_{\mbox{\tiny
eq}})&=&\sum_{\xi,\sigma}\Pi_0(\xi,\sigma)\ln[1+\epsilon(\xi,\sigma)]\\
&=&\sum_{\xi,\sigma}\Pi_{\mbox{\tiny eq}}(\xi,\sigma)
[1+\epsilon(\xi,\sigma)]\ln[1+\epsilon(\xi,\sigma)]\\
&=&\frac{1}{2}\sum_{\xi,\sigma}\Pi_{\mbox{\tiny
eq}}(\xi,\sigma)\epsilon^2(\xi,\sigma)+o(\epsilon^2)\\
&=&O(\epsilon^2)
\end{eqnarray}
and similarly, $D(\Pi_\tau\|\Pi_{\mbox{\tiny eq}})=O(\epsilon^4)$.
We have seen then that while both $kT\Delta\calH$ and
$\left<\Delta W\right>$ scale linearly with $\{\epsilon(\xi,\sigma)\}$ (for
small $\epsilon(\xi,\sigma)$),
the difference between them scales with $\{\epsilon^2(\xi,\sigma)\}$. Thus, if
both $\left<\Delta W\right>$ and $kT\Delta\calH$ are positive, the ratio
between them may be arbitrarily close to unity, provided that
$\{\epsilon(\xi,\sigma)\}$ are 
sufficiently small.

Even if $\Pi_0$ and $\Pi_{\mbox{\tiny eq}}$ differ considerably, it is still possible
to approach the entropy production bound, 
but this may require many small steps (in the spirit of quasi--static
processes in classical thermodynamics), 
i.e., a chain of many systems of the type of Fig.\ 1,
where the output bit--stream of each one of them serves as the 
input bit--stream to the next one. This 
approach was hinted already in
\cite{Merhav15} and later also in \cite{BMC16c}. If we think of $\Pi_0$ as the
canonical distribution with respect to some Hamiltonian $E_0(\xi,\sigma)$
(which is always possible, say, by defining $E_0(\xi,\sigma)=-kT\ln
\Pi_0(\xi,\sigma)$), then we can design a long sequence of distributions,
$\Pi^{(1)}, \Pi^{(2)}, \ldots, \Pi^{(L)}=\Pi_{\mbox{\tiny eq}}$ ($L$ -- large
positive integer), such that
$\Pi^{(i)}$ has
``Hamiltonian''
$(1-i/L)E_0(\xi,\sigma)+(i/L)E(\xi,\sigma)$, $i=1,2,\ldots,L$, so that the
distance between every two consecutive distributions (in the above sense) is of the order
of $\epsilon=1/L$ and hence the gap between the entropy production and the 
incremental work, pertaining to the
passage from $\Pi^{(i)}$ to $\Pi^{(i+1)}$, is of the order of $\epsilon^2=1/L^2$, so
that even if we sum up all these gaps, the total cumulative gap is of the order of
$L$ steps times $1/L^2$, which is $1/L$, and hence can still be made arbitrarily
small by selecting $L$ large enough.

\subsection{Memoryless and Markov Input Processes}

Most of the earlier works on systems with information reservoirs assumed that
the input process $\{X_n\}$ is memoryless,
i.e., that $P(x_1,\ldots,x_N)$ admits
a product form for all $N$. In this case, $S_n$, which is generated by
$X_1,\ldots,X_{n-1}$, must be statistically independent of $X_n$,
and so, in eq.\ (\ref{alternative}), $H(X_n|S_n)=H(X_n)$. We therefore obtain
from (\ref{alternative}), the following:
\begin{eqnarray}
\frac{\left<W_N\right>}{kT}&\le&\sum_{n=1}^N[H(Y_n|S_{n+1})-H(X_n)]+H(S_{N+1})-H(S_1)\\
&=&\sum_{n=1}^N[H(Y_n)-H(X_n]-\sum_{n=1}^NI(S_{n+1};Y_n)+H(S_{N+1})-H(S_1).
\end{eqnarray}
As already mentioned in the context of 
(\ref{alternative}), if we divide both sides by $N$ and take the limit
$N\to\infty$, the term $\frac{1}{N}[H(S_{N+1})-H(S_1)]\le\frac{1}{N}\ln|\calS|$ vanishes as
$N\to\infty$, and if we also drop the negative contribution of the mutual
information terms, we further enlarge the expression to
obtain the familiar bound that the asymptotic work per cycle cannot
exceed the limit of
$kT\cdot\frac{1}{N}\sum_{n=1}^N[H(Y_n)-H(X_n]$. As discussed also in \cite{BMC16a},
\cite{BMC16b}, \cite{BMC16c}, this bound is valid (and can be approached,
following the discussion in the previous subsection) also by a memoryless
ratchet, namely, a ratchet with one internal state only. Moreover, it is not
only that there is nothing to lose from using a memoryless ratchet, but on the
contrary -- there is, in fact, a lot to lose if the ratchet uses memory in a
non--trivial manner: this loss is
expressed in the negative 
term $-\sum_{n=1}^NI(S_{n+1};Y_n)$. The loss can, of course, be avoided if 
we make sure that at the
end of each cycle, the two
components of the Markov state, namely, $S_{n+1}$ and $Y_n$, are statistically
independent, and so, $I(S_{n+1};Y_n)=0$ for all $n$. If $\tau$ is large enough
so that $\Pi_{\mbox{\tiny eq}}$ is approached, and if $E(\xi,\sigma)$ is additive (namely,
$E(\xi,\sigma)=E_1(\xi)+E_2(\sigma)$), then $\Pi_{\mbox{\tiny eq}}(\xi,\sigma)=
\Pi_{\mbox{\tiny eq}}(\xi)\Pi_{\mbox{\tiny eq}}(\sigma)$, and this is the case. 
Indeed, in \cite{MJ12}, for example, this is the case, as there are six Markov states ($|\calX|=2$
times $|\calS|=3$) and $\Pi_{\mbox{\tiny
eq}}(\xi,\sigma)=e^{-\beta mgh\xi}/[3(1+e^{-\beta mgh})]$, $\xi\in\{0,1\},
\sigma\in\{A,B,C\}$.\\

\vspace{0.1cm}

\noindent
{\it Example.} Consider a binary memoryless source with
$\mbox{Pr}\{X_n=1\}=1-\mbox{Pr}\{X_n=0\}=p$, and a two--state ratchet, with a
state set
$\calS=\{A,B\}$. The joint process $\{(X_n,S_n)\}$ (as well as
$\{(\xi_t,\sigma_t)\}$ within each interaction interval) is therefore a
four--state process with state set $\{A0,B0,A1,B1\}$. Let the energy levels
be $E(A0)=0$, $E(B0)=\epsilon$, $E(A1)=2\epsilon$ and $E(B1)=3\epsilon$, where
$\epsilon > 0$ is a given energy quantum.
The Markov jump process $\{(\xi_t,\sigma_t)\}$ has transition rates,
$M[A0\to B0]=
M[B0\to A1]=
M[A1\to B1]=e^{-\beta\epsilon}$,
$M[B1\to A1]=
M[A1\to B0]=M[B0\to A0]=1$ (in some units of frequency) and 
all other transition rates are zero (see Fig.\ 2).
This process obeys detailed balance and its equilibrium distribution is given by
$\Pi_{\mbox{\tiny eq}}[A0]=1/Z$, $\Pi_{\mbox{\tiny eq}}[B0]=e^{-\beta\epsilon}/Z$,
$\Pi_{\mbox{\tiny eq}}[A1]=e^{-2\beta\epsilon}/Z$, and
$\Pi_{\mbox{\tiny eq}}[B1]=e^{-3\beta\epsilon}/Z$, where
$Z=1+e^{-\beta\epsilon}+e^{-2\beta\epsilon}+e^{-3\beta\epsilon}$.

\begin{figure}[ht]
\hspace*{1cm}\input{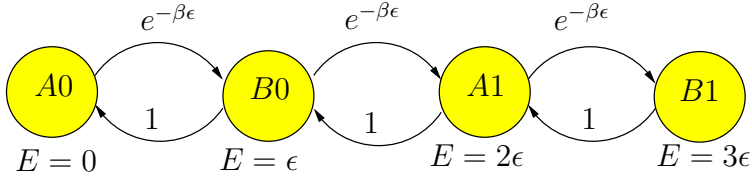}
\caption{Example of the Markov jump process.}
\end{figure}

Suppose that $\tau$ is very large compared to the time constants of the
process, so that $\Pi_\tau(\xi,\sigma)$ can be well approximated by the
equilibrium distribution. Then, it is straightforward to see
that
\begin{equation}
P(Y_n=0|S_{n+1}=A)=\frac{\Pi_{\mbox{\tiny eq}}[A0]}{\Pi_{\mbox{\tiny
eq}}[A0]+\Pi_{\mbox{\tiny eq}}[A1]}=\frac{1}{1+e^{-2\beta\epsilon}}
\end{equation}
and similarly for $P[Y_n=0|S_{n+1}=B]$. Therefore,
\begin{equation}
H(Y_n|S_{n+1})=h\left(\frac{1}{1+e^{-2\beta\epsilon}}\right),
\end{equation}
where $h(\cdot)$ is the binary entropy function, defined in Section 2.
As for the input entropy, we have $H(X_n|S_n)=H(X_n)=h(p)$. Therefore, the upper bound
on the work per cycle is
\begin{equation}
\left<\Delta W_n\right>\le h\left(\frac{1}{1+e^{-2\beta\epsilon}}\right)-h(p).
\end{equation}
It follows that a necessary condition for the ratchet to operate as an engine
(rather than as an eraser) is $p < 1/(1+e^{2\beta\epsilon})$ or
$p > 1/(1+e^{-2\beta\epsilon})$. Using similar considerations, the exact work extraction is also easy to
calculate in this example, but we will not delve into it any further.
This concludes the example.

Consider next the case where the input process is a stationary first order Markov
process, i.e.,
\begin{equation}
P(x^N)=P(x_1)\prod_{n=1}^{N-1}P(x_{n+1}|x_n).
\end{equation}
As described above, in the discrete time scale, 
the ratchet is characterized by the input--output transition probability
distribution
$P(y,s^\prime|x,s)=\mbox{Pr}\{Y_n=y,S_{n+1}=s^\prime|X_n=x,S_n=s\}$. 
Consider the corresponding marginal conditional
distribution
\begin{equation}
P(s^\prime|x,s)=\sum_{y\in\calX}P(y,s^\prime|x,s).
\end{equation}
Then, assuming that the initial ratchet state, $S_1$, is independent of the
initial input symbol, $X_1$, we have
\begin{equation}
P(x^N,s^N)=P(x_1)P(s_1)\prod_{n=1}^{N-1}[P(x_{n+1}|x_n)P(s_{n+1}|x_n,s_n)],
\end{equation}
which means that the pair process $\{(X_n,S_n)\}$ is a first order Markov
process as well. Let us assume that the transition matrix of this Markov pair
process is such that there exists a unique stationary distribution
$P(x,s)=\mbox{Pr}\{X_n=x,S_n=s\}$. Once the stationary distribution $P(x,s)$
is found, the input--output--state joint distribution is dictated by the
ratchet input--output transition probability distribution
$\{P(y,s^\prime|x,s)\}$, according to
\begin{equation}
P(x,s,y,s^\prime)=P(x,s)P(y,s^\prime|x,s),
\end{equation}
which is the joint distribution of the quadruple $(X_n,S_n,Y_n,S_{n+1})$ in
the stationary regime. Once this joint distribution is found, one can (relatively) easily compute the 
stationary average work extraction per cycle,
$\left<\Delta W_n\right>=\left<E(Y_n,S_{n+1})\right>-\left<E(X_n,S_n)\right>$,
as well as the stationary joint
entropies $H(X_n,S_n)$ and $H(Y_n,S_{n+1})$ (or $H(X_n|S_n)$ and
$H(Y_n|S_{n+1})$) in order to calculate the entropy--production bound.
This should be contrasted with the bound in \cite{BMC16a} (see also
\cite{BMC16b}, \cite{BMC16c}), where, as mentioned earlier, $\left<W_N/NkT\right>$ is asymptotically
upper bounded by $\lim_{N\to\infty}\frac{1}{N}[H(Y^N)-H(X^N)]$, whose
calculation is not trivial, as $Y^N$ is a hidden Markov process, for which
there is no closed--form expression for the entropy rate. 

A good design of a
ratchet would be in the quest of finding the transition distribution
$\{P(y,s^\prime|x,s)\}$ that maximizes the work extraction (or its entropy
production bound) for the given Markov input process. This is an optimization
problem with a finite (and fixed) number of parameters.
If, in addition, one has the freedom to 
control the parameters of the Markov input process, say, by transducing a
given source of randomness, e.g., a random bit--stream, then of course, 
the optimization will include also the
induced joint distribution $\{P(x,s)\}$. If such a transducer is a
one--to--one mapping, then its operation does not consume energy.
For example, if the raw input stream is a sequence of independent fair coin
tosses (i.e., a purely random bit--stream), this transducer can be chosen to
be the decoder of an optimal lossless data
compression scheme for the desired input process $P$.

\subsection{Conditional Entropy Bounds}

We now return to the case of a general input process.
For a given $n=1,2,\ldots$, let us denote $u_n=(x^{n-1},y^{n-1},s^n)$, which
is the full
input--output--state history available at time $n$, and define $v_n=f_n(u_n)$,
where $f_n$ is an arbitrary function. If $f_n$ is a many--to--one function, then
$v_n$ designates some partial history information, for example, $v_n=x^{n-1}$,
or $v_n=y^{n-1}$. Once again, when we wish to emphasize the randomness of all
these variables, we use capital letters: $U_n=(X^{n-1},Y^{n-1},S^n)$,
$V_n=f_n(U_n)$, etc.
Now consider the application of the H--theorem (eq.\
(\ref{htheorem})) with $\Pi_0(\xi,\sigma)= P(X_n=\xi,S_n=\sigma|V_n=v_n)$, instead of the unconditional
distribution as before. Then, using the Markovity of the
dynamics within each interaction interval, the same derivation as in Section 3
would now yield
\begin{eqnarray}
\left<\Delta W_n|V_n=v_n\right>&\equiv&\left<E(Y_n,S_{n+1})|V_n=v_n\right>-
\left<E(X_n,S_n)|V_n=v_n\right>\nonumber\\
&\le&
kT[H(Y_n,S_{n+1}|V_n=v_n)-H(X_n,S_n|V_n=v_n)],
\end{eqnarray}
where the notation $\left<\cdot|V_n=v_n\right>$ designates conditional
expectation given $V_n=v_n$. Averaging both sides with respect to (w.r.t.) the randomness of
$V_n$, we get
\begin{eqnarray}
\left<\Delta W_n\right>&\equiv&\left<E(Y_n,S_{n+1})\right>-
\left<E(X_n,S_n)\right>\nonumber\\
&\le&
kT[H(Y_n,S_{n+1}|V_n)-H(X_n,S_n|V_n)],
\end{eqnarray}
and summing all inequalities from $n=1$ to $n=N$, we obtain the family of
bounds,
\begin{eqnarray}
\label{conditional}
\left<W_N\right>&\equiv&\sum_{n=1}^N[\left<E(Y_n,S_{n+1})\right>-
\left<E(X_n,S_n)\right>]\nonumber\\
&\le&
kT\sum_{n=1}^N[H(Y_n,S_{n+1}|V_n)-H(X_n,S_n|V_n)],
\end{eqnarray}
with a freedom in the choice 
of $V_n$ (or, equivalently, the
choice of the function $f_n$).
Now, one may wonder what is the best choice
that would yield
the tightest bound in this family. Conditioning reduces entropy, but it
reduces both the entropy of $(Y_n,S_{n+1})$ and that of $(X_n,S_n)$, so it may
not be immediately clear what happens to the difference.
A little thought, however, shows that the best choice of $V_n$ is
null, namely, the unconditional entropy bound of Section 3 is no worse than
any bound of the form (\ref{conditional}). To see why this is true, observe
that
\begin{eqnarray}
& &H(Y_n,S_{n+1}|V_n)-H(X_n,S_n|V_n)\nonumber\\
&=&H(Y_n,S_{n+1})-H(X_n,S_n)+I(V_n;X_n,S_n)-I(V_n;Y_n,S_{n+1})\\
&\ge&H(Y_n,S_{n+1})-H(X_n,S_n),
\end{eqnarray}
where the inequality follows from the data processing inequality
\cite[Sect.\ 2.8]{CT06}, as $V_n$ and $(Y_n,S_{n+1})$ are statistically independent
given $(X_n,S_n)$, owing to the Markov property of the process
$\{(\xi_t,\sigma_t)\}$. Consequently, $I(V_n;X_n,S_n)\ge I(V_n;Y_n,S_{n+1})$,
and the inequality is achieved when $V_n$ is degenerate.
Thus, for the purpose of upper bounding the work, the conditioning on any
partial history $V_n$ turns out to be completely useless.

However, the family of inequalities (\ref{conditional}) may be more
interesting when we
consider them as lower bounds on entropy production rather
than upper bounds on extractable work. Specifically, consider the case
$V_n=(X^{n-1},Y^{n-1})$. Then, the work/entropy-production inequality reads
\begin{eqnarray}
\label{entropybound}
\frac{\left<W_N\right>}{kT}&\le&
\sum_{n=1}^N[H(Y_n,S_{n+1}|X^{n-1},Y^{n-1})-H(X_n,S_n|X^{n-1},Y^{n-1})]\nonumber\\
&\le&\sum_{n=1}^N[H(Y_n,S_{n+1}|Y^{n-1})-H(X_n|X^{n-1},Y^{n-1})-\nonumber\\
& &H(S_n|X^n,Y^{n-1})]\nonumber\\
&=&\sum_{n=1}^N[H(Y_n|Y^{n-1})+H(S_{n+1}|Y^n)-\nonumber\\
& &H(X_n|X^{n-1})-H(S_n|X^n,Y^{n-1})]\nonumber\\
&=&H(Y^N)-H(X^N)+\sum_{n=1}^N[H(S_{n+1}|Y^n)-H(S_n|X^n,Y^{n-1})],
\end{eqnarray}
where the first equality is since $Y^{n-1}$ is independent of
$X_n$ given $X^{n-1}$. Now, the second term in the last line
of eq.\ (\ref{entropybound}) is equivalent to
\begin{eqnarray}
& &H(S_{N+1}|Y^N)-H(S_1|X_1)+\sum_{n=2}^N[H(S_n|Y^{n-1})-H(S_n|X^n,Y^{n-1})]\nonumber\\
&=&H(S_{N+1}|Y^N)-H(S_1|X_1)+\sum_{n=2}^NI(S_n;X^n|Y^{n-1}).
\end{eqnarray}
In general, this expression can always be upper bounded by $N\ln|\calS|$, and
so, we obtain the following lower bound on the output entropy
\begin{equation}
H(Y^N)\ge H(X^N)+\frac{\left<W_N\right>}{kT}-N\ln|\calS|.
\end{equation}
Suppose now that $\{X_n\}$ is a memoryless process, or even a Markov process.
Then, as mentioned earlier (see also \cite{BMC16a,BMC16b,BMC16c}), 
$\{Y_n\}$ is a hidden Markov process, and as already explained before, the joint entropy of $Y^N$
is difficult to compute and it does not have a simple closed--form expression.
On the other hand, the above lower bound on $H(Y^N)$ is relatively easy to calculate, as
$P(x^N)$ has a simple product form and $\left<W_N\right>$ depends only on the
marginals of $(X_n,S_n)$ and $(Y_n,S_{n+1})$, which can be calculated
recursively from the transition probabilities $\{P(y_n,s_{n+1}|x_n,s_n)\}$,
for $n=1,2,\ldots,N$, and if in addition, $\{X_n\}$ is stationary, then
$(X_n,Y_n)$ and $(Y_n,S_{n+1})$ have stationary distributions too, as
described before.
While one may suspect that $N\ln|\calS|$ might be a loose bound for the second
term on the right--most side of (\ref{entropybound}), there are, nevertheless,
situations where it is quite a reasonable bound, especially when $\ln|\calS|$ is small
(compared to $H(X^N)/N+\left<W_N\right>/NkT$). Moreover, if the marginal entropy of $S_n$
is known to be upper bounded by some constant $H_0 < \ln|\calS|$, then $\ln|\calS|$ can be
replaced by $H_0$ in the
above lower bound.

\section{More General Inequalities}

An equivalent form of the basic result of Section 3 is the following:
\begin{equation}
\calH(\Pi_0)-\beta\left<E(\xi_0,\sigma_0)\right>\le
\calH(\Pi_\tau)-\beta\left<E(\xi_\tau,\sigma_\tau)\right>,
\end{equation}
The l.h.s.\ can be thought of as the negative free energy of the
Markov state at time $t=0$
(multiplied by a factor of $\beta$),
and the r.h.s.\ is the same quantity at time $t=\tau$.
In other words, if we define the random variable 
\begin{equation}
\phi_t(\xi,\sigma)=-\ln
\Pi_t(\xi,\sigma)-\beta E(\xi,\sigma), 
\end{equation}
then what we have seen in Section
3 is that 
\begin{equation}
\left<\phi_0(\xi,\sigma)\right>_0\le
\left<\phi_t(\xi,\sigma)\right>_t, 
\end{equation}
where $\left<\cdot\right>_t$ denotes expectation w.r.t.\ $\Pi_t$.
Equivalently, if we denote $\phi(X_n,S_n)=-\ln P(X_n,S_n)-\beta E(X_n,S_n)$,
$\phi(Y_n,S_{n+1})=-\ln P(Y_n,S_{n+1})-\beta E(Y_n,S_{n+1})$,
and we take $t=\tau$, this becomes
\begin{equation}
\label{phinequality}
\left<\phi(X_n,S_n)\right>\le
\left<\phi(Y_n,S_{n+1})\right>, 
\end{equation}
where the expectations at both sides are w.r.t.\ the randomness of the
relevant random variables.

In this section, we show that this form of the inequality relation extends to
more general moments of the random variables $\phi(X_n,S_n)$ and
$\phi(Y_n,S_{n+1})$. As is well known, the H--theorem applies to generalized
divergence functionals and not only to the Kullback--Leibler divergence
$D(\Pi_t\|\Pi_{\mbox{\tiny eq}})$, see
\cite[Theorem 1.6]{Kelly79}, \cite[Chap.\ V.5]{vanKampen}.
Let $Q$ be any convex function and suppose that $\Pi_{\mbox{\tiny eq}}(x,s)> 0$
for every $(x,s)$. Then according to the generalized H--theorem,
\begin{equation}
D_Q(\Pi_t\|\Pi_{\mbox{\tiny eq}})=\sum_{x,s} \Pi_{\mbox{\tiny
eq}}(x,s)Q\left(\frac{\Pi_t(x,s)}{\Pi_{\mbox{\tiny eq}}(x,s)}\right)
\end{equation}
decreases monotonically as a function of $t$, and so,
\begin{equation}
D_Q(\Pi_\tau\|\Pi_{\mbox{\tiny eq}})\le 
D_Q(\Pi_0\|\Pi_{\mbox{\tiny eq}}).
\end{equation}
Now, 
\begin{eqnarray}
D_Q(\Pi_t\|\Pi_{\mbox{\tiny eq}})&=&\left<\frac{\Pi_{\mbox{\tiny
eq}}(\xi,\sigma)}{\Pi_t(\xi,\sigma)}\cdot Q\left(\frac{\Pi_\tau(\xi,\sigma)}{P{\mbox{\tiny
eq}}(\xi,\sigma)}\right)\right>_t\nonumber\\
&=&\frac{1}{Z}\cdot\left<e^{\phi_t(\xi,\sigma)}
\cdot Q\left(Z\cdot e^{-\phi_t(\xi,\sigma)}\right)\right>_t
\end{eqnarray}
In the corresponding inequality between $D_Q(\Pi_\tau\|\Pi_{\mbox{\tiny eq}})$ and
$D_Q(\Pi_0\|\Pi_{\mbox{\tiny eq}})$, the external factor
of $1/Z$, obviously cancels out. Also, since $Q(u)$ is convex iff $Q(Z\cdot
u)$ ($Z$ -- constant) is convex, we can re--define the latter as our convex function $Q$
to begin with, and so, by the generalized H--theorem\footnote{Note that the
classical H--theorem
is obtained as a special case by the choice $Q(u)=u\ln u$.}
\begin{equation}
\Lambda(t)\equiv\left<e^{\phi_t(\xi,\sigma)}\cdot Q\left(
e^{-\phi_t(\xi,\sigma)}\right)\right>_t
\end{equation}
is monotonically decreasing for any convex function $Q$.
It now follows that
\begin{equation}
\left<e^{\phi(X_n,S_n)}\cdot Q\left(e^{-\phi(X_n,S_n)}\right)\right>\ge
\left<e^{\phi(Y_n,S_{n+1})}\cdot Q\left(e^{-\phi(Y_n,S_{n+1})}\right)\right>.
\end{equation}
This class of inequalities has the flavor of fluctuation theorems concerning
$\phi(X_n,S_n)$ and $\phi(Y_n,S_{n+1})$.
We observe that unlike the classical H--theorem, which makes a claim only about the
the first
moments of $\phi(X_n,S_n)$ and $\phi(Y_n,S_{n+1})$, here we have a more general statement concerning
the monotonicity of moments of a considerably wide family of functions of
these random variables.
For example, choosing $Q(u)=-\ln u$ gives 
\begin{equation}
\left<\phi(X_n,S_n)e^{\phi(X_n,S_n)}\right>
\ge\left<\phi(Y_n,S_{n+1})e^{\phi(Y_n,S_{n+1})}\right>,
\end{equation}
which is somewhat counter--intuitive, in view of (\ref{phinequality}),
as the function $f(u)=ue^u$ is monotonically increasing.

An interesting family of functions $\{Q\}$ is the family of power functions,
defined as $Q_z(u)=u^{1-z}$ for $z\le 0$ and $z\ge 1$
and $Q_z(u)=-u^{1-z}$ for $z\in[0,1]$.
Here we obtain that 
\begin{equation}
\left<\exp\{z\phi(X_n,S_n)\}\right>\le
\left<\exp\{z\phi(Y_n,S_{n+1})\}\right>~~~~~~~~~\mbox{for}~~z\in[0,1]
\end{equation}
and
\begin{equation}
\left<\exp\{z\phi(X_n,S_n)\}\right>\ge
\left<\exp\{z\phi(Y_n,S_{n+1})\}\right>~~~~~~~~~\mbox{for}~~z\notin[0,1]
\end{equation}
Note that for $z > 1$, $P(X_n=x,S_n=s)$
must be strictly positive for all $(x,s)$ with $E(x,s) < \infty$, for otherwise,
there is a singularity. We have therefore obtained inequalities that involve
the characteristic functions of $\phi(X_n,S_n)$ and $\phi(Y_n,S_{n+1})$. It is
interesting to observe that the direction of the inequality is reversed when
the parameter $z$ crosses both the values $z=0$ and $z=1$.

\section*{Appendix A}

\noindent
{\it Some Concerns About the Derivation of Eq.\ (5) of
\cite{BMC16a}.}

In Appendix A of \cite{BMC16a}, eq.\ (5) of that paper is derived, namely, the
inequality that upper bounds the work extraction per cycle by $kT$ times the
difference between the Shannon entropy rate of the output process and that of the
input process, as mentioned in the Introduction. 
The derivation in
\cite[Appendix A]{BMC16a} begins from the second law of thermodynamics, and on
the basis of the second law, it states that the joint {\it Shannon entropy} of the
entire system, consisting of the ratchet state, the input tape, the output
tape, and the heat bath, must not decrease with time (eq.\ (A2) in
\cite{BMC16a}). 

The first concern is that
while the second law is an assertion about the increase of the thermodynamic
entropy (which is, strictly speaking, defined for equilibrium), some more 
care should be exercised when addressing the
increase of the Shannon entropy. 
To be specific, we are familiar with two situations (in classical statistical
physics) where
the Shannon entropy is known to be non--decreasing. The first is associated with
Hamiltonian dynamics, where the total Shannon entropy simply remains fixed, due
to the Liouville theorem, as
argued, for example, in \cite[Section III]{DJ13}, and indeed, ref.\
\cite{DJ13} is cited in \cite{BMC16a} (in the context of eq.\ (2) therein),
but there is no assumption in \cite{BMC16a} about Hamiltonian dynamics, and it
is not even clear that Hamiltonian dynamics can be assumed in this model
setting, in the first place, due to the discrete nature of the input and output information
streams, as well as the ratchet state.
Two additional assumptions made in \cite{DJ13}, but not in \cite{BMC16a}, are that the
system is initially prepared in a product state (i.e., the states of the
different parts of the system are statistically independent) \cite[eq.\
(27)]{DJ13} and that the heat
bath is initially in equilibrium 
\cite[eq.\ (28)]{DJ13}. By contrast, the only assumption
made in \cite{BMC16a} is that the ratchet has a finite number of states
(see first sentence in \cite[Appendix A]{BMC16a}).

The second situation where the Shannon entropy is known 
to be non--decreasing is when the
state of the system is a Markov process, which has a
uniform stationary state distribution, owing to the H--Theorem (see, for example,
\cite[Chap.\ V, Sect.\ 5]{vanKampen}). However, it is not clear that the
total system under discussion obeys Markov dynamics with a stationary
distribution (let alone, the uniform distribution), 
because the tape moves in one direction only, so states
accessible at a given time instant are no longer accessible at later times
(after $n$ cycles, the machine has converted $n$ input bits to output bits,
so the position of the tape relative to the ratchet, indexed by $n$, 
should be part of the Markovian state).

Another concern is that in Appendix A of \cite{BMC16a}, it is argued that the
state of the heat bath is independent of the states of the ratchet and the
tape at all times, with the somewhat vague 
explanation that ``they have no memory of the environment''
(see the text immediately after eq.\ (A4) of \cite{BMC16a}). 
While this independence argument may make sense 
with regard to the initial preparation (at time $t=0$) of the
system (again, as assumed also in \cite{DJ13}), it is less clear why this
remains true also at later times, after the systems have interacted for a 
while. Note that indeed, in
\cite{DJ13}, the various components of the system are not assumed
independent at positive times.

To summarize, there seems to be some room for concern that more
assumptions may be needed in \cite{BMC16a} beyond the assumption on a finite
number of ratchet states.

\end{document}